# GdV$_6$Sn$_6$: a Multi-carrier Metal with Non-magnetic 3*d*-electron Kagome Bands and 4*f*-electron Magnetism

Hajime Ishikawa*, Takeshi Yajima, Mitsuaki Kawamura,
Hiroyuki Mitamura, and Koichi Kindo

[1]*Institute for Solid State Physics, the University of Tokyo, 5-1-5 Kashiwanoha, Kashiwa, Chiba, Japan*

Electronic properties of the single crystal of GdV$_6$Sn$_6$, where non-magnetic V-kagome layers are separated by magnetic Gd-triangular lattice, are investigated. GdV$_6$Sn$_6$ exhibits unique magnetotransport properties at low-temperature such as non-linear Hall resistivity and increase of resistance $R$ in magnetic field $H$ as $R \sim H^{0.75}$ up to 56 T with Shubnikov–De Haas oscillations. Investigation of the non-magnetic analogue YV$_6$Sn$_6$ and the first principles calculations reveal these properties are relevant to the bands arising from the V-kagome layer. A magnetic transition at 5 K in GdV$_6$Sn$_6$ modifies the transport properties, pointing to a coupling between Gd-spins on the triangular lattice and carriers in the V-kagome layer.

## 1. Introduction

Kagome lattice, a two-dimensional network made of corner shared triangles, has attracted condensed matter physicists as a platform for realizing novel electronic states. A localized spin system on the kagome lattice is a typical example that exhibits a quantum spin liquid ground state due to the geometrical frustration [1,2]. Itinerant electrons on the kagome lattice produces a band structure with uncommon features such as a Dirac point, van Hove singularity, and a flat band and may host unconventional density wave orders or superconductivity [3-5].

Experimentally, 3*d* transition metal compounds with kagome lattice have been investigated in search for novel correlated electronic phases. Kagome cuprates are candidates for the Mott insulator with a quantum spin liquid ground state [6,7,8]. Intermetallic compounds with the kagome layers of Cr, Mn, Fe, and Co have been investigated in recent years as a metal with large anomalous Hall effect [9] and novel band structures [10-12]. More recently, a series of vanadium antimonide $A$V$_3$Sb$_5$ ($A$ = K, Rb, Cs) have attracted considerable attention as a new member of the kagome metal [13]. $A$V$_3$Sb$_5$ compounds exhibit various phenomena such as the co-existence of charge density wave (CDW) order and superconductivity [14] and the anomalous Hall effect without a magnetic order [15,16], suggesting that vanadium kagome metal is a promising platform for investigating unconventional electronic properties.

Kagome layers made of vanadium are found in a series of stannide $R$V$_6$Sn$_6$ ($R$ = Rare Earth), which are discovered in 2011 in the polycrystalline form [17]. Their physical properties are not investigated in the first report. The short V-V distances of approximately 2.7 Å may produce metallic conductivity within the kagome layer as seen in $A$V$_3$Sb$_5$. The variety of the rare earth ions, which separate the kagome layers instead of alkaline metals in $A$V$_3$Sb$_5$, may allow us to tune the physical properties. In this study, we have synthesized the single crystals of GdV$_6$Sn$_6$ and YV$_6$Sn$_6$ and investigate their electronic properties. Based on the experimental results and first principles calculations, we argue that GdV$_6$Sn$_6$ and YV$_6$Sn$_6$ are multi-carrier metal with band structure arising from the 3$d$-electron in the kagome layers, which exhibit peculiar magnetotransport properties. A coupling of Gd-spins forming a triangular lattice and conduction carriers are found in GdV$_6$Sn$_6$ at low temperature and low magnetic field region.

## 2. Experimental Results

High-quality single crystals of GdV$_6$Sn$_6$ and YV$_6$Sn$_6$, which are supported by the good convergence of the structural analyses of the single crystal x-ray diffraction data, are prepared by the flux method. The polycrystalline samples are also prepared and used to measure magnetic susceptibility at high temperature: see supplemental material for the details of experimental methods and crystallographic information [18]. The crystal structure of the single crystals of GdV$_6$Sn$_6$ and YV$_6$Sn$_6$ are essentially consistent with the previous powder sample [17] (Fig.1, depicted by VESTA software [19]). There are two kagome layers made of V in a unit cell with the V-V distance of 2.7701(2) and 2.7634(3) Å within the kagome plane in GdV$_6$Sn$_6$ and YV$_6$Sn$_6$, respectively, which are very close to each other and also similar to those found in the $A$V$_3$Sb$_5$ family [13]. The kagome layers are separated by Gd/Y atoms forming a triangular lattice and Sn atoms. There are three crystallographically distinct Sn sites: the site disorder discussed in the first report [17] was not detected in our structural analysis on the single crystals.

The temperature dependence of magnetic susceptibility $M/H(T)$ per formula unit of the polycrystalline sample of YV$_6$Sn$_6$ exhibits almost constant value of $1 \times 10^{-3}$ emu/mol in the wide temperature range, suggesting the compound is a non-magnetic metal (Fig. 2(a)). A fit by the Curie-Weiss law $M/H(T) = \chi_0 + C/(T + \Theta)$, where $\chi_0$, $C$ and $\Theta$ indicate the temperature independent term, Curie's constant, and Weiss-temperature respectively, yields $\chi_0 = 1.05(5) \times$

$10^{-3}$ emu/mol, $C = 8.5(2) \times 10^{-4}$ emu/mol K, and $\Theta = 2.4(1)$ K. The $C$ correspond to 0.8% of the spin-1/2 per V atom, indicating the upturn at low temperature comes from a tiny amount of certain impurity phase. $M/H(T)$ of GdV$_6$Sn$_6$ is much larger than that of YV$_6$Sn$_6$. The Curie-Weiss fit above 10 K yields $\chi_0 = 1.9(2) \times 10^{-3}$ emu/mol, $C = 7.86(2)$ emu/mol K, and $\Theta = -3.83(2)$ K. The $C$ corresponds to the effective magnetic moment $\mu_{eff} = 7.93$ $\mu_B$, which is very close to 7.94 $\mu_B$ calculated for the Gd$^{3+}$ ($4f^7$) ion with $S = 7/2$. $\Theta$ indicates the net magnetic interaction is weakly ferromagnetic.

Magnetic properties of GdV$_6$Sn$_6$ at low temperatures are investigated in detail using a single crystal (Fig. 2(b)). $M/H(T)$ measured in the magnetic field perpendicular to the kagome plane ($H // c$) at 0.01 T exhibits a kink at 5 K, suggesting a magnetic transition. More pronounced kink is observed in the measurements in the magnetic field parallel to the kagome plane ($H // a$). There is a small hysteresis in the increasing and decreasing temperature processes at 0.01 T only when the magnetic field is applied along the $c$-axis, which becomes smaller at 0.1 T and absent at 0.2 T. Such behaviors may be caused by the geometrical frustration and/or competing interactions on a triangular lattice, which may stabilize complex magnetic orders [20]. The hysteresis in $H // c$ is also visible in the magnetization curve at 1.8 K: see supplemental material for the magnetization curve [18]. The anomaly in $M/H(T)$ is suppressed at 1 T. The magnetization almost amounts to the full moment of 7 $\mu_B$ at 1 T, which is consistent with the net ferromagnetic interactions.

The electric resistivity $\rho(T)$ of GdV$_6$Sn$_6$ and YV$_6$Sn$_6$ decreases as lowering the temperature as expected for a metal (Fig. 3(a)). $\rho$ is as low as a few $\mu\Omega$cm at low temperature, indicating the high electric conductivity. The residual resistivity ratio $\rho(320\ \text{K})/\rho(1.8\ \text{K})$ is 26 and 38 in GdV$_6$Sn$_6$ and YV$_6$Sn$_6$, respectively, indicating the high quality of the crystals. Anomaly suggestive of the CDW order found in the $A$V$_3$Sb$_5$ family are not observed. $\rho(T)$ of GdV$_6$Sn$_6$ exhibits a kink at 5 K, where the magnetic transition is observed in $M/H(T)$, suggesting a coupling between the Gd-spins and the conduction carriers.

GdV$_6$Sn$_6$ and YV$_6$Sn$_6$ exhibit relatively large changes of the in-plane longitudinal resistivity $\rho_{xx}$ at low temperature in the magnetic field perpendicular to the kagome plane (Fig. 3(b)). GdV$_6$Sn$_6$ exhibits negative magnetoresistance, $(\rho_{xx}(H) - \rho_{xx}(0\ \text{T}))/\rho_{xx}(0\ \text{T})$, of -12% at 1 T and 1.8 K (Fig. 3(c)). On the other hand, non-magnetic YV$_6$Sn$_6$ exhibits positive magnetoresistance, suggesting the negative magnetoresistance is caused by the coupling between the magnetism and conduction carriers. Indeed, GdV$_6$Sn$_6$ exhibits the positive magnetoresistance at 20 K above the magnetic transition temperature.

Above 1 T, where spins are almost fully polarized, the magnetoresistance increases and reaches to +25% in GdV$_6$Sn$_6$ at 10 T and 1.8 K. The resistance $R$ of a crystal of GdV$_6$Sn$_6$ is measured up to 56 T in the pulsed high magnetic field at 1.4 K. $R$ increases monotonically up to 20 T and Shubnikov–De Haas oscillations appear at higher fields. The oscillating component $\Delta R$ is obtained by subtracting the background, which increases approximately as $R \sim H^{0.75}$ (Fig.4). The $R \sim H^{0.75}$ behavior is also observed in YV$_6$Sn$_6$ at 1.8 K and GdV$_6$Sn$_6$ at 20 K at lower field region (dashed lines in Fig.3(c)), suggesting it is attributed to the common V-kagome layer.

The in-plane Hall resistivity $\rho_{xy}$ of GdV$_6$Sn$_6$ and YV$_6$Sn$_6$ exhibit similarly complex temperature evolution, suggesting the presence of multiple carriers (Fig.4(a)). The $\rho_{xy}$ is linear and positive at around room temperature, indicating the dominant carrier is electron. The carrier number estimated assuming single carrier at 320 K is 2.4 and 3.2 × 10$^{22}$ cm$^{-3}$ in GdV$_6$Sn$_6$ and YV$_6$Sn$_6$, respectively. The non-linear behavior of the Hall resistivity appears as lowering the temperature and the slope at high field region becomes negative, suggesting the contribution from holes appear at low temperature. At 1.8 K, $\rho_{xy}$ exhibits clear $S$-shaped feature at low-field region below 1 T and behaves linearly at higher fields. Similar non-linear Hall resistivity is reported in the $A$V$_3$Sb$_5$ family [15,16].

## 3. Discussion

To elucidate the electronic properties of GdV$_6$Sn$_6$ and YV$_6$Sn$_6$, we performed the first principles calculations: see supplemental material for the calculation method and the detailed band structure plot [18]. The Fermi surface of YV$_6$Sn$_6$ plotted by using the FermiSurfer program [21] indicates the presence of cylinder with large cross-section area around the Γ-point and small pockets between the L- and M- points (Fig.5(a)). Tiny Fermi surfaces are found at the K-point, where Dirac points appear in the tight binding model on the kagome lattice [3]. There are nodal lines between M points. The calculation for GdV$_6$Sn$_6$ is performed including the effect of spin splitting (Fig.5(b)). The obtained band structure should mimic the band structure in the strong magnetic field. The overall band structure is unchanged but the shapes and sizes of the pockets, Dirac points, and nodal lines are modified.

The Shubnikov–De Haas oscillations at high-fields can be reproduced by the calculation using the Lifshitz-Kosevich formula with an oscillation frequency of approximately 180 T. The fitting can be significantly improved by assuming two components with oscillation frequencies of approximately 150 and 200 T: see supplementary information for details of the fit [18]. The frequencies correspond to the cross-section area of approximately 2% of the first

Brillouin zone and indicates the presence of two small Fermi surfaces with a similar size. The calculated size of the extremal cross-section area of the pocket between L- and M- points is 2.9 % in $YV_6Sn_6$ of the first Brillouin zone. The pockets are modified in $GdV_6Sn_6$ by the spin splitting and takes the sizes of 2.0-4.6% of the first Brillouin zone. The sizes of the pockets exhibit relatively good agreement with the observed quantum oscillation. Therefore, it would be plausible to attribute the observed quantum oscillations to the small pockets between L- and M- points.

The orbital projected density of states obtained from the first principles calculation indicate the dominant V-3$d$ contribution at the Fermi level in $GdV_6Sn_6$ and $YV_6Sn_6$ (Fig.5(c,d)), indicating the transport properties of the compounds are dominated by the V-kagome layer. This is consistent with the similar temperature dependence and $S$-shaped behavior of Hall resistivity observed commonly in $GdV_6Sn_6$ and $YV_6Sn_6$. When the linear part above 1 T is subtracted from the Hall resistivity at 1.8 K, the remaining feature $\Delta\rho_{xy}$ (Fig.4(b)) bears striking similarity to those observed in the $AV_3Sb_5$ compounds [15,16]. Some mechanism such as scattering related to the spin clusters [15] or the CDW order [16] on the kagome lattice are proposed as the origin of the non-linear Hall behavior. Such scenarios are not compatible with $YV_6Sn_6$ which hosts neither magnetism nor CDW order.

We performed the fits of the Hall resistivity of $GdV_6Sn_6$ and $YV_6Sn_6$ at 1.8 K by the two-carrier model. The simultaneous refinement of the four parameters, carrier numbers of electron and hole ($n_e$ and $n_h$) and their mobilities ($\mu_e$ and $\mu_h$), did not converge. Here, tentatively, we estimated $n_h$ from the linear fit at high field region and fix its mobility to a certain value, and then refine the parameters of the electron. The fits assuming $\mu_h$ of 1000 cm$^2$/V·s yields reasonably good fits for the non-linear Hall resistivity both in $GdV_6Sn_6$ and $YV_6Sn_6$ (Fig4(a)). The fitting suggests that the presence of multiple carriers with high mobility may reproduce the observed non-linear Hall resistivity at low temperature. Note that our results do not exclude the possibility of an anomalous Hall effect caused by a certain mechanism.

While the low-temperature Hall resistivity are seemingly reproduced by the two-carrier model, the $R \sim H^{0.75}$ behavior observed both in $GdV_6Sn_6$ and $YV_6Sn_6$ in the wide magnetic field range are not compatible with the conventional two-carrier model, which predicts $H^2$ behavior at low field region and saturation at high field region. In some Dirac semimetals, linearly increasing magnetoresistance is reported [22]. The $H^{0.75}$ behavior, which is close to linear, may be relevant to the presence of the Dirac crossings at the K-point.

## 4. Conclusion

We have synthesized the high-quality single crystals of two isostructural vanadium kagome metals $GdV_6Sn_6$ and $YV_6Sn_6$ and investigated their electronic properties. The compounds are multi-carrier metal with unique magnetotransport properties such as non-linear Hall resistivity and quasi-linear magnetoresistance. Small Fermi pockets, Dirac points, and nodal lines found in the first principles calculation may be relevant to such properties. Magnetic transition at 5 K observed in $GdV_6Sn_6$ modifies the resistivity, indicating a coupling between Gd-spins and conduction carriers. Our results call for future experiments such as determination of the magnetic structure and examination of the magnetic field angle dependence of the quantum oscillations and Hall resistivity to understand the unusual transport properties. At the same time, our results demonstrate that $RV_6Sn_6$ family provides a new class of material to investigate the interplay between itinerant non-magnetic 3$d$-electron with bands inherent to the kagome lattice and magnetism of the 4$f$-electron on the triangular lattice.

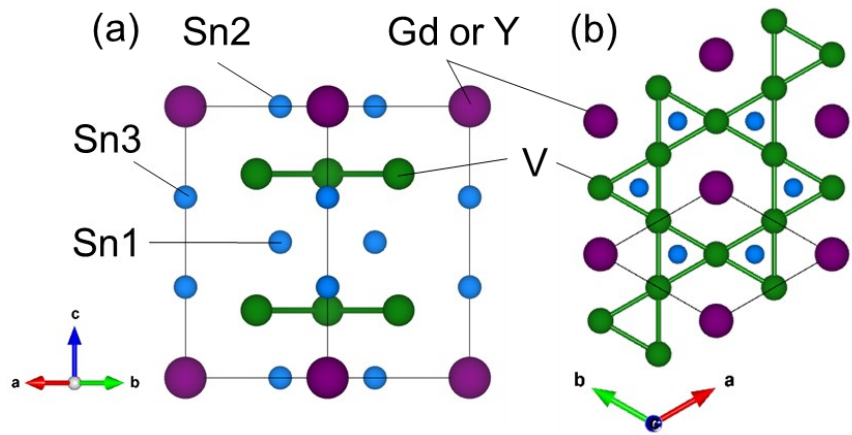

Fig. 1. (Color online) Crystal structure of $GdV_6Sn_6$ and $YV_6Sn_6$ determined by the single crystal x-ray diffraction measurements: see supplemental material and the cif-file for details of the crystallographic data [18].

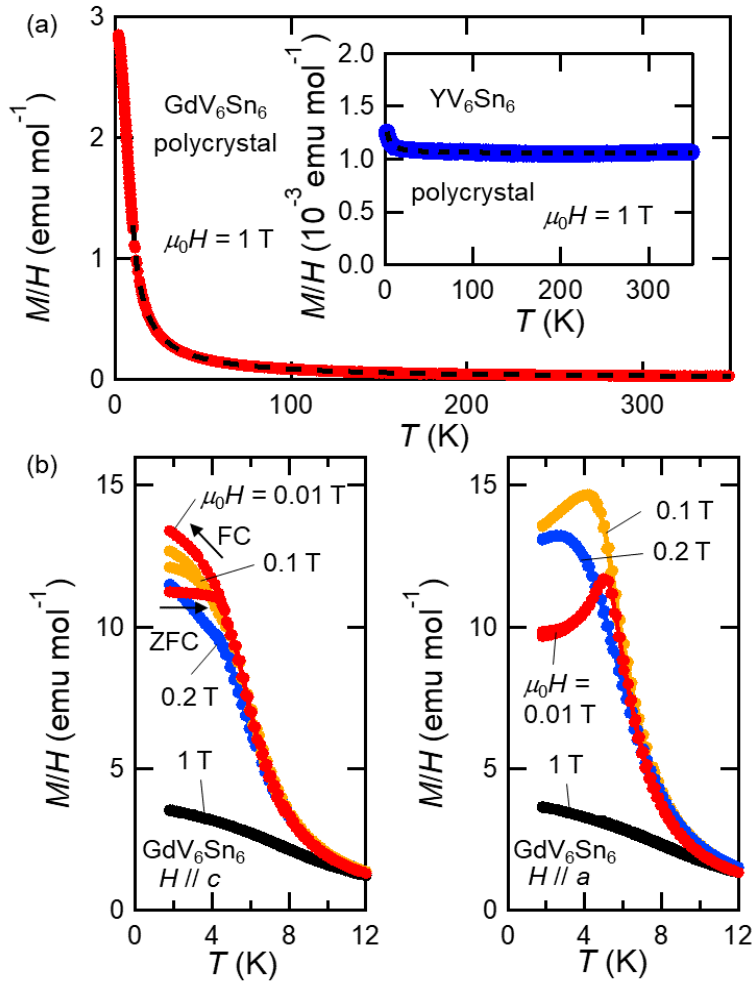

Fig. 2. (Color online) Temperature dependence of the magnetic susceptibility of (a) polycrystalline samples of GdV$_6$Sn$_6$ and YV$_6$Sn$_6$ and (b) a single crystal of GdV$_6$Sn$_6$ measured in $H \mathbin{/\mkern-2mu/} c$ and $a$. The dashed lines in (a) indicate the Curie-Weiss fit. Data in the zero-field cooling (ZFC) and field cooling (FC) measurements are shown for all the magnetic field values and clear hysteresis is visible in the data with $H = 0.01$ and $0.1$ T applied along the $c$-axis.

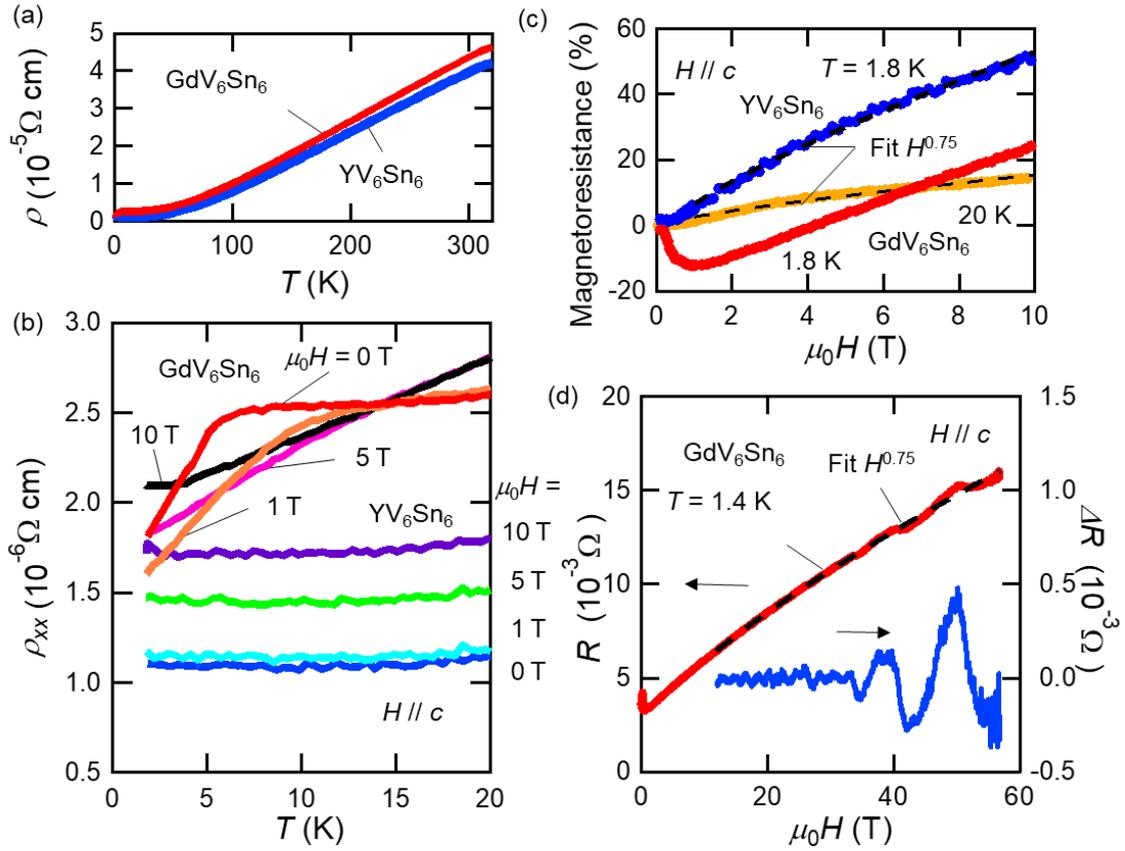

Fig. 3 (Color online) (a) Temperature dependence of in-plane resistivity of the single crystals of GdV$_6$Sn$_6$ and YV$_6$Sn$_6$ between 1.8 and 320 K (b) In-plane resistivity between 1.8 and 20 K in the magnetic field applied along the $c$-axis. (c) Magnetoresistance of GdV$_6$Sn$_6$ at 1.8 K and 20 K and that of YV$_6$Sn$_6$ at 1.8 K up to 10 T. A fit to $H^{0.75}$ magnetic field dependence is shown for the latter two data by the dashed lines. (d) Magnetoresistance of GdV$_6$Sn$_6$ at 1.4 K measured up to 56 T in pulsed magnetic field. A fit to $H^{0.75}$ magnetic field dependence, which is used to estimate the oscillating component $\Delta R$, is shown by the dashed line.

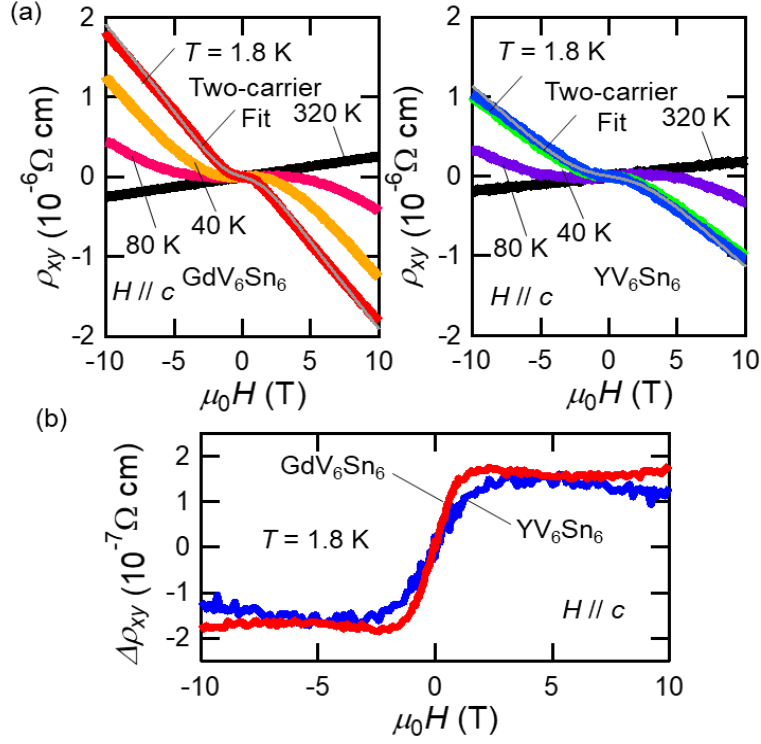

Fig. 4 (Color online) (a) In-plane Hall resistivity of $GdV_6Sn_6$ and $YV_6Sn_6$ measured at different temperatures in $H // c$. An examples of the fit by the two-carrier model is shown by the solid line for the data at 1.8 K: fits by the parameters $n_e = 1.2 \times 10^{20}$ cm$^{-3}$, $n_h = 3.2 \times 10^{21}$ cm$^{-3}$, $\mu_e = 4000$ cm$^2$/V·s, and $\mu_h = 1000$ cm$^2$/V·s for $GdV_6Sn_6$ and $n_e = 4.8 \times 10^{20}$ cm$^{-3}$, $n_h = 5.0 \times 10^{21}$ cm$^{-3}$, $\mu_e = 2500$ cm$^2$/V·s, and $\mu_h = 1000$ cm$^2$/V·s for $YV_6Sn_6$ are sown. (b) The non-linear component of the Hall resistivity of $GdV_6Sn_6$ and $YV_6Sn_6$ at 1.8 K.

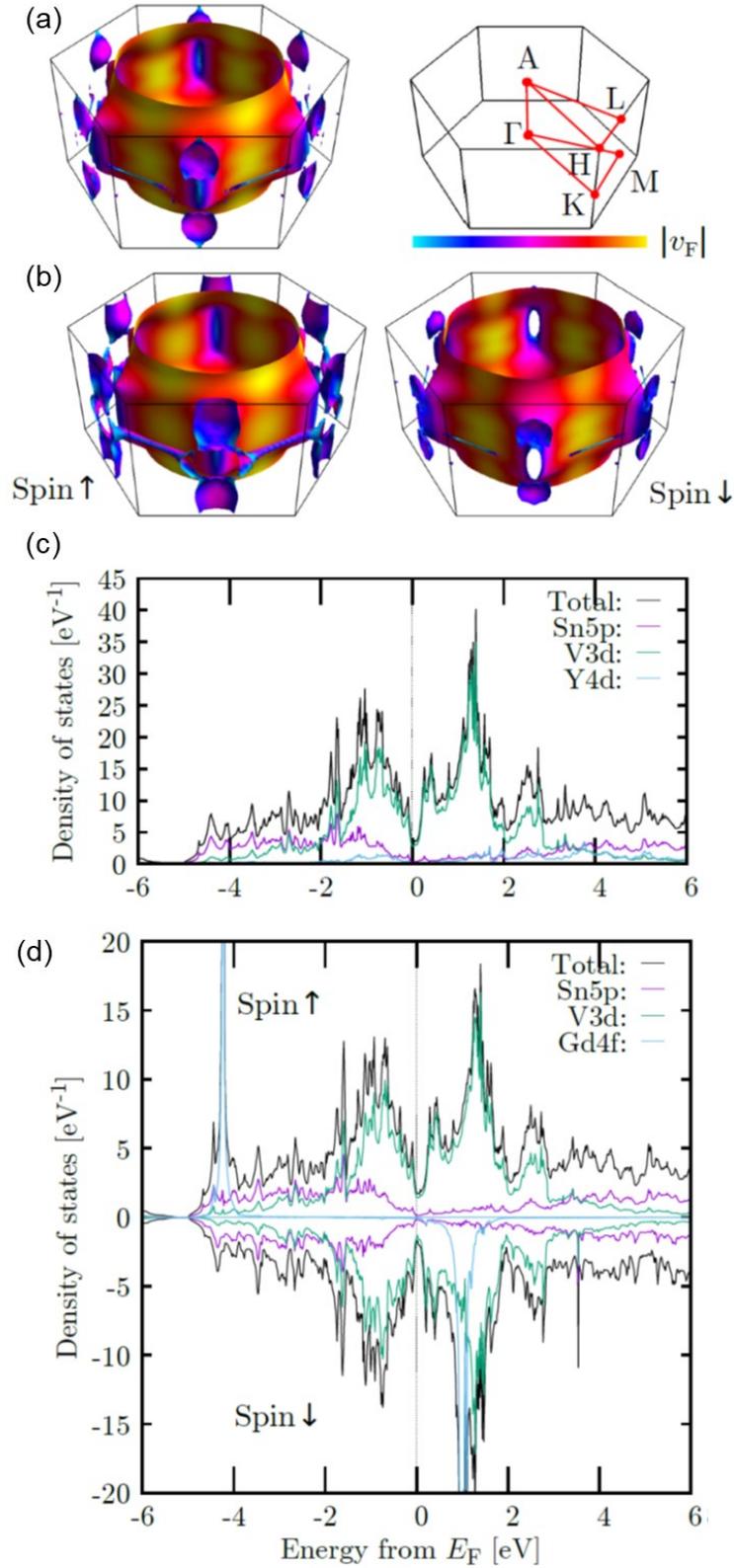

Fig. 5 (Color online) (a) Fermi surface plot of YV$_6$Sn$_6$ and the first Brillouin zone. (b) Fermi surface plot of GdV$_6$Sn$_6$ including the effect of spin splitting. Orbital projected density of states of (c) YV$_6$Sn$_6$ and (d) GdV$_6$Sn$_6$.

**References**

1) S. Yan, D. A. Huse, and S. R. White, *Science*, 332, 1173 (2011)

2) S. Depenbrock, I. P. McCulloch, and U. Schollwöck, *Phys. Rev. Lett.* 109, 067201 (2012).

3) H.-M. Guo and M. Franz, *Phys. Rev. B*, 80, 113102 (2009).

4) M. L. Kiesel, C. Platt, and R. Thomale, Phys. Rev. Lett. 110, 126405 (2013).

5) I. I. Mazin, H. O. Jeschke, F. Lechermann, H. Lee, M. Fink, R. Thomale and R. Valentí, Nat. Commun. 5, 4261 (2014).

6) Z. Hiroi, M. Hanawa, N. Kobayashi, M. Nohara, H. Takagi, Y. Kato, and M. Takigawa, J. Phys. Soc. Jpn. 70, 3377 (2001).

7) M. P. Shores, E. A. Nytko, B. M. Bartlett, and D. G. Nocera, J. Am. Chem. Soc. 127, 13462 (2005).

8) Y. Okamoto, H. Yoshida, and Z. Hiroi J. Phys. Soc. Jpn. 78, 033701 (2009).

9) S. Nakatsuji, N. Kiyohara, and T. Higo, Nature 527.7577 (2015): 212-215.

10) Y. Ishii, H. Harima, Y. Okamoto, J. Yamaura, Z. Hiroi, J. Phys. Soc. Jpn., 82, 023705 (2013).

11) N. Morali, R. Batabyal, P. K. Nag, E. Liu, Q. Xu, Y. Sun, B. Yan, C. Felser, N. Avraham, H. Beidenkopf, Science, 365, 1286 (2019).

12) M. Kang, L. Ye, S. Fang, J.-S. You, A. Levitan, M. Han, J. I. Facio, C. Jozwiak, A. Bostwick, E. Rotenberg, M. K. Chan, R. D. McDonald, D. Graf, K. Kaznatcheev, E. Vescovo, D. C. Bell, E. Kaxiras, J. van den Brink, M. Richter, M. P. Ghimire, J. G. Checkelsky, and R.Comin, Nat. Mater., 19, 163 (2020).

13) B. R. Ortiz, L. C. Gomes, J. R. Morey, M. Winiarski, M. Bordelon, J. S. Mangum, I. W. H. Oswald, J. A. Rodriguez-Rivera, J. R. Neilson, St. D. Wilson, E. Ertekin, T. M. McQueen, E. S. Toberer, Phys. Rev. Materials 3, 094407 (2019).

14) B. R. Ortiz, S. M. L. Teicher, Y. Hu, J. L. Zuo, P. M. Sarte, E. C. Schueller, A. M. M. Abeykoon, M. J. Krogstad, S. Rosenkranz, R. Osborn, R. Seshadri, L. Balents, J. He, S. D. Wilson, Phys. Rev. Lett. 125, 247002 (2020).

15) S.-Y. Yang, Y. Wang, B. R. Ortiz, D. Liu, J. Gayles, E. Derunova, R. Gonzalez-Hernandez, L. Šmejkal, Y. Chen, S. S. P. Parkin, S. D. Wilson, E. S. Toberer, T. McQueen, M. N. Ali, Sci. Adv., 6, eabb6003 (2020).

16) F. H. Yu, T. Wu, Z. Y. Wang, B. Lei, W. Z. Zhuo, J. J. Ying, and X. H. Chen, Phys. Rev. B 104, L041103 (2021).

17) L. Romaka, Y. Stadnyk, V. V. Romaka, P. Demchenko, M. Stadnyshyn, M. Konyk, J. alloys compd., 509, 8862 (2011).

18) (Supplemental material) Details of the experimental and calculation methods, crystallographic data, magnetization curve of GdV$_6$Sn$_6$, analysis of the quantum oscillation, and band structure plots are provided online. The cif-files of GdV$_6$Sn$_6$ and YV$_6$Sn$_6$ are provided separately.
19) K. Momma, F. Izumi, J. Appl. Crystallogr., 44, 1272-1276 (2011).
20) S. Hayami, R. Ozawa, and Y. Motome, Phys. Rev. B 95, 224424 (2017).
21) M. Kawamura, Comp. Phys. Commun 239, 197 (2019).
22) J. Feng, Y. Pang, D. Wu, Z. Wang, H. Weng, J. Li, X. Dai, Z. Fang, Y. Shi, and L. Lu Phys. Rev. B 92, 081306(R) (2015).

*E-mail: hishikawa@issp.u-tokyo.ac.jp

# Supplemental Material for
# GdV$_6$Sn$_6$: a Multi-carrier Metal with Non-magnetic 3$d$-electron Kagome Bands and 4$f$-electron Magnetism

Hajime Ishikawa*, Takeshi Yajima, Mitsuaki Kawamura,
Hiroyuki Mitamura, and Koichi Kindo

$^1$Institute for Solid State Physics, the University of Tokyo,
5-1-5 Kashiwanoha, Kashiwa, Chiba, Japan

## 1. Experimental and Calculation Methods

Single crystals of GdV$_6$Sn$_6$ and YV$_6$Sn$_6$ are prepared by the flux method. Gd or Y pieces (99.9%), V powder (99.5%), and Sn pieces (99.999%) are weighed with the molar ratio of 1 : 6 : 40 and put in an alumina crucible. The crucible is sealed in an evacuated quarts glass tube. The tube is heated up to 1020°C and then cooled slowly typically by 2°C/h. The tube is removed from the furnace at around 600°C and excess Sn is removed by centrifuge. Hexagonal plate-shaped single crystals of $R$V$_6$Sn$_6$ are found at the bottom of the crucible with certain polycrystalline materials. Impurities on the surface of the crystals such as residual Sn flux are removed by soaking the crystals into diluted hydrochloric acid. Single crystal x-ray diffraction (XRD) measurements are performed by a diffractometer with Mo$K\alpha$1 radiation (RIGAKU, RAPID). Polycrystalline samples of GdV$_6$Sn$_6$ and YV$_6$Sn$_6$, which are used in the magnetic susceptibility measurements at high temperatures, are prepared as previously reported [S1]. The stoichiometric amounts of elements are melted by the arc furnace under argon atmosphere. The reaction product is sealed in an evacuated quarts glass tube and annealed at 800°C for 20 days. The phase purity of the sample is confirmed by the powder XRD measurement using a diffractometer with Cu K$\alpha$ radiation (RIGAKU, SmartLab).

The electric resistivity and Hall resistivity are measured for the single crystals at 1.8-320 K in the magnetic field up to 10 T by the commercial apparatus (PPMS, Quantum Design). The electric resistivity of the single crystal of GdV$_6$Sn$_6$ is measured by an AC method [S2] up to 56 T in pulsed magnetic field generated at the International MegaGauss Science Laboratory at ISSP, the University of Tokyo. The magnetic susceptibility is measured by the SQUID magnetometer (MPMS, Quantum Design).

The first-principles calculation based on density functional theory is performed by using the Quantum ESPRESSO program package [S3], which employs plane waves and pseudopotentials to describe the Kohn-Sham orbitals and the crystalline potential, respectively.

The plane-wave cutoff for a wave function is set to 60 Ry. We employ the GGA-PBE functional [S4]. The ultrasoft pseudopotential [S5] (for Y, V, and Sn) and the dataset of projector-augmented wave (PAW) [S6] (for Gd) used in this study are obtained in Standard Solid State Pseudopotentials [S7]. They are included in the Garrity-Bennett-Rabe-Vanderbilt pseudopotential library [S8] and the Rare Earth PAW datasets [S9]. We set the $k$-point grids of Brillouin-zone integrations with the optimized tetrahedron method [S10] for the charge density and the Fermi surface to 8×8×4 and 40×40×30, respectively. For the calculation of GdV$_6$Sn$_6$, we consider the ferromagnetic spin configuration possible under a high magnetic field.

## 2. Structural Parameter of GdV$_6$Sn$_6$ and YV$_6$Sn$_6$ determined by single crystal X-ray diffraction experiments

**GdV$_6$Sn$_6$**

Space Group: Hexagonal $P6/mmm$, $a$ = 5.5403(3) Å, $c$ = 9.2046(6) Å

| Atom | x   | y   | z           | occupancy | 100$U_{iso}$ |
|------|-----|-----|-------------|-----------|--------------|
| Gd   | 0   | 0   | 0           | 1         | 0.67(2)      |
| V    | 0   | 1/2 | 0.24855(13) | 1         | 0.47(3)      |
| Sn1  | 2/3 | 1/3 | 1/2         | 1         | 0.56(2)      |
| Sn2  | 2/3 | 1/3 | 0           | 1         | 0.47(2)      |
| Sn3  | 0   | 0   | 0.33437(10) | 1         | 0.60(2)      |

$R_1$ = 0.0147 for 138 $F$o > 4σ($F$o) and 0.0166 for all 151 data

**YV$_6$Sn$_6$**

Space Group: Hexagonal $P6/mmm$, $a$ = 5.5267(4) Å, $c$ = 9.1907(6) Å

| Atom | x   | y   | z          | occupancy | $U_{iso}$ |
|------|-----|-----|------------|-----------|-----------|
| Y    | 0   | 0   | 0          | 1         | 0.72(3)   |
| V    | 0   | 1/2 | 0.24834(8) | 1         | 0.47(2)   |
| Sn1  | 2/3 | 1/3 | 1/2        | 1         | 0.57(2)   |
| Sn2  | 2/3 | 1/3 | 0          | 1         | 0.47(2)   |
| Sn3  | 0   | 0   | 0.33349(6) | 1         | 0.62(2)   |

$R_1$ = 0.0140 for 146 $F$o > 4σ($F$o) and 0.0151 for all 151 data

More detailed crystallographic information is available in the attached cif-files.

### 3. Magnetization curve of GdV$_6$Sn$_6$

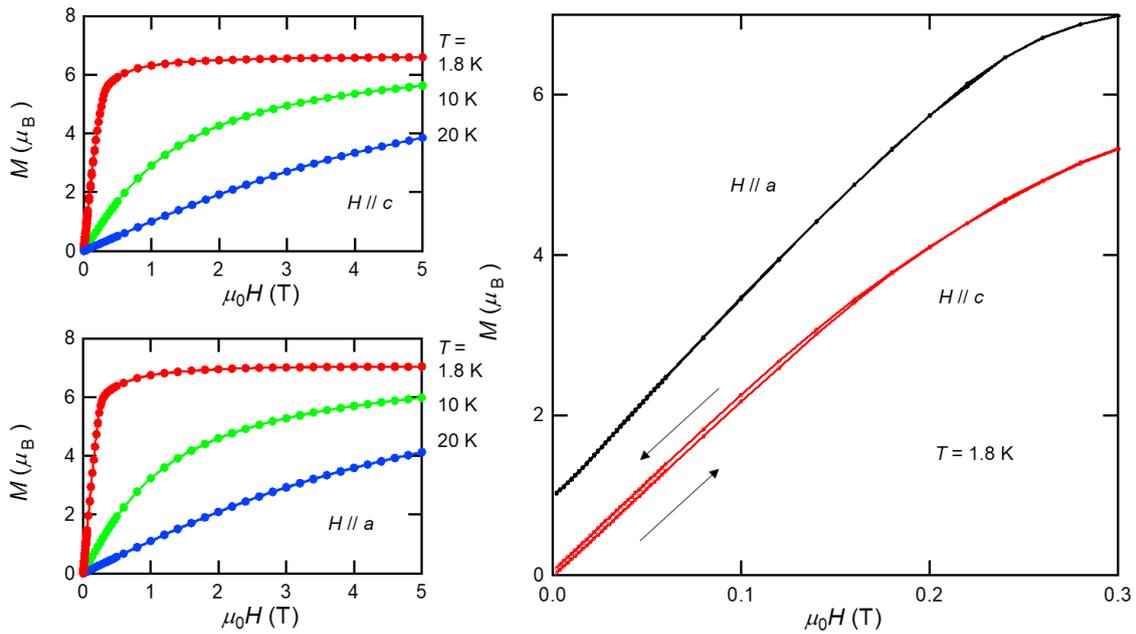

The magnetization curves of GdV$_6$Sn$_6$ at different temperatures (left). The magnetization curve in $H \parallel c$ exhibits a small hysteresis below 0.2 T in the increasing and decreasing magnetic fields (right). Such hysteresis is not observed in $H \parallel a$ (plotted with an offset of 1 $\mu_B$): data in the increasing and decreasing processes overlap with each other.

### 4. Fits of the Shubnikov–De Haas oscillations

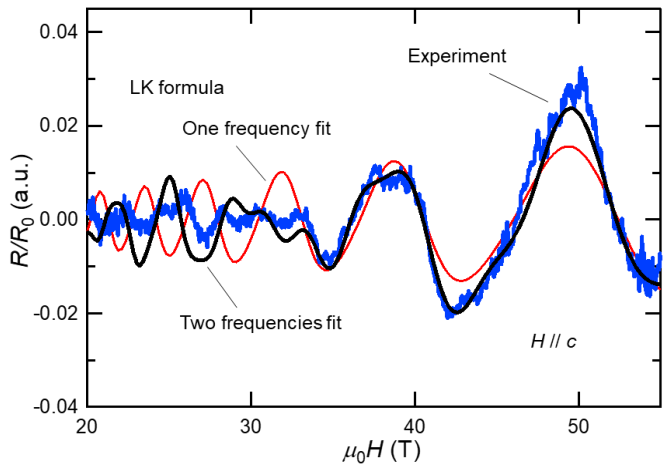

The experimentally observed quantum oscillation (blue) at the high-magnetic region can be reproduced by a fit with the Lifshitz-Kosevich (LK) formula [S11] assuming oscillation frequency of 180 T (red). The fit is significantly improved by assuming two components with the oscillation frequencies of 150 and 200 T (black), suggesting the presence of two Fermi surfaces with a similar size.

## 5. Band structure plots of YV$_6$Sn$_6$ and GdV$_6$Sn$_6$

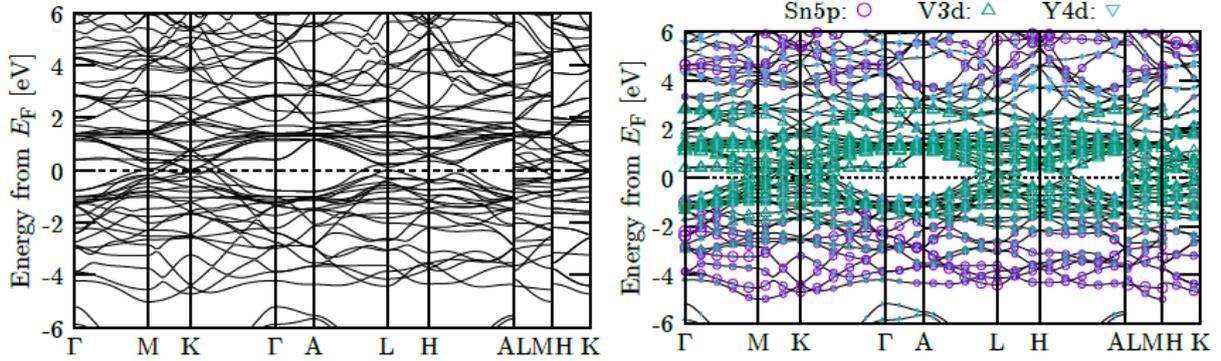
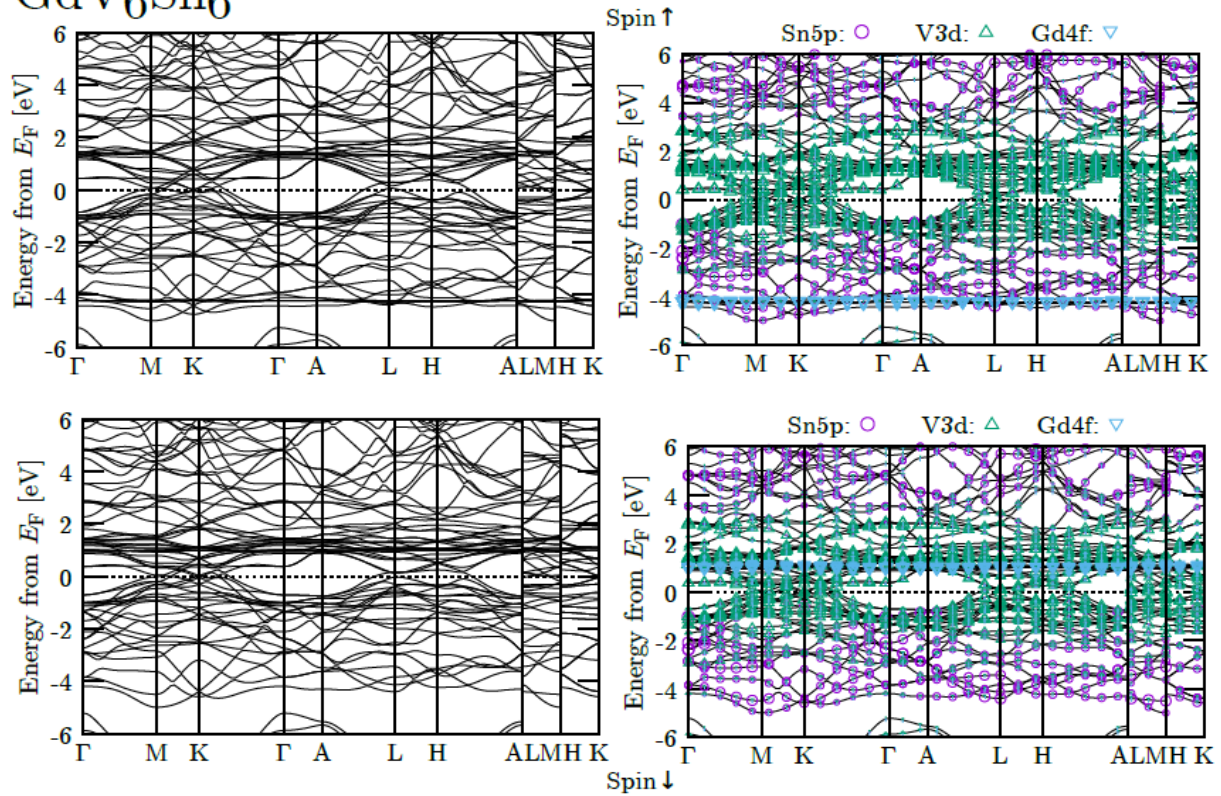

## References

[S1] L. Romaka, Y. Stadnyk, V. V. Romaka, P. Demchenko, M. Stadnyshyn, M. Konyk, J. Alloys Compd., 509, 8862 (2011).

[S2] H. Mitamura, R. Watanuki, E. Kampert, T. Förster, A. Matsuo, T. Onimaru, N. Onozaki, Y. Amou, K. Wakiya, K. T. Matsumoto, I. Yamamoto, K. Suzuki, S. Zherlitsyn, J. Wosnitza, M. Tokunaga, K. Kindo, T. Sakakibara, Rev. Sci. Instrum. 91, 125107 (2020).

[S3] P. Giannozzi, O. Andreussi, T. Brumme, O. Bunau, M. Buongiorno Nardelli, M. Calandra, R. Car, C. Cavazzoni, D. Ceresoli, M. Cococcioni, N. Colonna, I. Carnimeo, A. Dal Corso, S. de Gironcoli, P. Delugas, R. A. DiStasio Jr, A. Ferretti, A. Floris, G. Fratesi, G. Fugallo, R. Gebauer, U. Gerstmann, F. Giustino, T Gorni, J Jia, M. Kawamura, H.-Y. Ko, A. Kokalj, E. Küçükbenli, M. Lazzeri, M. Marsili,

N. Marzari, F. Mauri, N L Nguyen, H.-V. Nguyen, A. Otero-de-la-Roza, L. Paulatto, S. Poncé, D. Rocca, R. Sabatini, B. Santra, M. Schlipf, A. P. Seitsonen, A. Smogunov, I. Timrov, T. Thonhauser, P. Umari, N. Vast, X. Wu and S. Baroni, J. Phys.: Condens. Matter 29, 465901 (2017).

[S4] J. P. Perdew, K. Burke, and M. Ernzerhof, Phys. Rev. Lett. 78, 1396 (1997).

[S5] D. Vanderbilt, Phys. Rev. B 41, 7892(R) (1990).

[S6] P. E. Blöchl, Phys. Rev. B 50, 17953 (1994).

[S7] G. Prandini, A. Marrazzo, I. E. Castelli, N. Mounet, and N. Marzari, npj Comput. Mater. 4, 72 (2018).

[S8] K. F. Garrity, J. W. Bennett, K. M. Rabe, and D. Vanderbilt, Comput. Mater. Sci. 81, 446 (2014).

[S9] M. Topsakal and R. M. Wentzcovitch, Comput. Mater. Sci. 95, 263 (2014).

[S10] M. Kawamura, Y. Gohda, and S. Tsuneyuki, Phys. Rev. B 89, 094515 (2014).

[S11] D. Shoenberg, Magnetic oscillations in metals (Cambridge University Press, Cambridge, 1984).

*E-mail: hishikawa@issp.u-tokyo.ac.jp